\documentclass[prl, aps, amssymb, amsmath, showpacs, preprint]{revtex4}
\usepackage{epsf}
\usepackage{amssymb}
\usepackage{amsmath}
\begin{document}

\title{ Superluminal telecommunication: an observable contradiction
between quantum entanglement and relativistic causality}
\author{Ruo Peng WANG}
\email{rpwang@cis.pku.edu.cn} \affiliation{ Physics Department and
State Key Laboratory for mesoscopic Physics, Peking University,
Beijing 100871, P.R.China }
\date{\today}

\begin{abstract}
I present a schema for a superluminal telecommunication system based
on polarization entangled photon pairs. Binary signals can be
transmitted at superluminal speed in this system, if entangled
photon pairs can really be produced.  The existence of the
polarization entangled photon pairs is in direct contradiction to
the relativistic causality in this telecommunication system. This
contradiction implies the impossibility of generating entangled
photon pairs.

\end{abstract}

\pacs{03.65.Ud, 42.50.Dv}
\maketitle

Quantum non-locality \cite{epr} is a controversial topic of quantum
theory. Due to its quantum nature, it is generally believed that the
quantum non-local correlation between entangled particles does not
produce observable non-local effect, such as superluminal
telecommunication, that could contradict the relativistic causality.
The schema of the quantum telecommunication is based on the quantum
teleportation \cite{tele} is often used to justify this affirmation.
In that schema a classical channel \cite{telt} must be used,
therefore no superluminal telecommunication could be realized in
such a telecommunication system, in spite of the fact that this
system is based on the quantum non-local correlation between
entangled photon pairs. As a matter of fact, if one tries to
associate binary values to different quantum states, it will be
found that it is not possible to encode any information because the
result of a measurement is unpredictable,  and it is not possible to
decode  any information because a quantum state can not be
determined by a single measurement.

But however, one can encode and decode information in another way.
In this letter I present a schema for a telecommunication system
using polarization entangled photon pairs as signal carrier in which
the superluminal telecommunication can be realized, and the
relativistic causality is in direct contradiction with the existence
of entangled photon pairs.

The suggested superluminal telecommunication system based on
entangled photon pairs is schematically illustrated in Figs. 1, 2
and 3. This system consists of three components: a source of
polarization entangled photon pairs, a signal encoder, and a signal
decoder. Binary signals are transmitted in this telecommunication
system.

During telecommunication processes, polarization entangled photon
pairs are generated at the source S. One of the two entangled
photons is sent to the signal sender, and the other one is sent to
the signal receiver. The signal encoder placed at the signal sender
include a mirror (M) which can be switched between the position
labeled ``0'' and  the position labeled ``1'', and two photon
detector channels: the channel ``0'' and the channel ``1''. There
are one Glan-Thompsom prism and two single photon detectors in each
channel. The Glan-Thompsom prism (BS$_0$) in the channel ``0'' is
arranged in such a way that the photons detected by the photon
detector D$_0$ are horizontally polarized, the photons detected by
the photon detector D$'_0$ are vertically polarized. The
Glan-Thompsom prism (BS$_1$) in the channel ``1'' is rotated
$45^\circ$, so the photons detected by the photon detector D$'_1$
are polarized at $45^\circ$ to horizontal direction, and photons
detected by the photon detector D$'_1$ are polarized at $-45^\circ$
to horizontal direction.

\begin{figure}
\begin{center}
\epsfbox{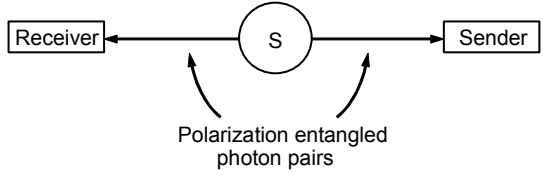}
\end{center}
\caption{\label{fig1} A schematic show of a superluminal
telecommunication based on entangled photon pairs. S is a source of
polarization entangled photon pairs}
\end{figure}

\begin{figure}
\begin{center}
\epsfbox{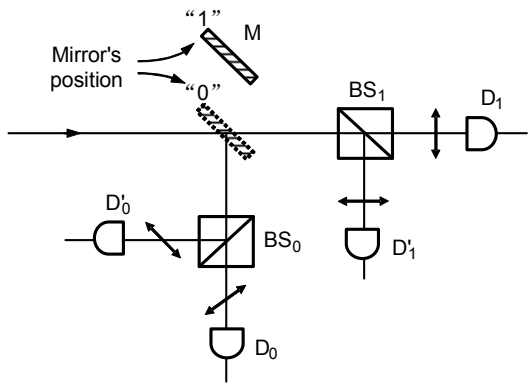}
\end{center}
\caption{\label{fig2} A schematic illustration of a signal encoder.
BS$_0$ and BS$_1$ are Glan-Thompsom prisms. D$_0$, D$_1$, D$'_0$,
D$'_1$, are single photon detectors. Polarizations of optical beams
are indicated by arrows.}
\end{figure}

\begin{figure}
\begin{center}
\epsfbox{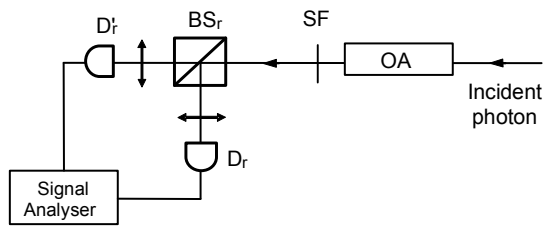}
\end{center}
\caption{\label{fig3} A schematic illustration of a signal decoder.
OA is an optical amplifier. BS$_r$ is a Glan-Thompsom prism. D$_r$,
D$'_r$ are single photon detectors. Polarizations of optical beams
are indicated by arrows.}
\end{figure}

The signal receiver is equipped with the signal decoder that
includes an optical amplifier (OA), a spatial filter (SF) that
allows only photons in the same spatial mode as the incident photon
to pass, a Glan-Thompsom prism (BS$_r$) placed at the output end of
the optical amplifier which splits an optical beam into horizontally
polarized and vertically polarized secondary beams. Two single
photon detectors (D$_r$ and D$'_r$) are placed in each of these two
secondary beams. The photon counts by the detectors D$_r$ and D$'_r$
are analysed by the signal analyser, and converted to the signal
readout.

When a photon entered in the OA, stimulated and spontaneous
emissions both take place. As just one photon is present, the
stimulated emitting rate equals the spontaneous emitting rate for
photons with the same polarization as the incident photon. At the
same time, photons with polarization perpendicular to the
polarization incident photon are also generated, at the same
spontaneous emitting rate. So the optical beam that passes the
spatial filter (SF) contains $2m+1$ photons with the same
polarization as the incident photon, and $m$ photons with
perpendicular polarization. In other words, an incident photon is
amplified by OA into a partial polarized optical beam. The
characteristic of this partial polarized optical beam depends on the
polarization of the incident photon.

To send a binary signal ``0'', the sender switches the mirror M into
the position ``0''. In this case the photon sent to the sender can
be detected either by D$_0$ or by D$'_0$. Once the photon being
detected, the polarization of the photon sent to the receiver is
also determined: it must be either horizontally polarized or
vertically polarized. The whole system can be setup in such a way
that the polarization of the photon sent to the receiver is
determined when it arrives at the receiver.  This ``signal carrier''
photon passes the optical amplifier OA, where $2m$ more photons with
the same polarization as the ``signal carrier'' photon and $m$
photon with perpendicular polarization are generated. In the case of
sending the signal ``0'', $2m+1$ photons with the same polarization
as the ``signal carrier'' photon are detected by one of the
detectors D$_r$ and D$'_r$, and $m$ photons with perpendicular
polarization are detected by another detector. A difference of $m+1$
in photon counts by the detectors D$_r$ and D$'_r$ corresponds to a
binary signal ``0'' at readout, and the signal ``0'' is received in
this way.

To send a binary signal ``1'', the sender switches the mirror M into
the position ``1''. In this case, in optical beam filtered by SF,
$2m+1$ photons are either polarized at $45^\circ$ or $-45^\circ$ to
horizontal direction, $m$ photons are either polarized at
$-45^\circ$ or $45^\circ$ to horizontal direction. Thus each photon
can be detected either by the detector D$_r$ or by the detector
D$'_r$ with equal probability. The average photon counts difference
is null in this case. A a null average photon counts difference
corresponds to a binary signal ``1'' at readout, and the signal
``1'' is received. Evidently, we must take the photon counts
fluctuation into consideration. The signal to noise rate in photon
counts is proportional to $\sqrt{m}$, thus for large $m$, the signal
``0'' and signal ``1'' can be clearly separated.

In the case of the signal ``1'', there exist a non-zero probability
for finding a difference of photon counts close to $m+1$. If this
happens, then instead the corrected signal ``1'', an erroneous
signal ``0'' is read out. Although this error is inevitable, the
probability for such an error to happen can be reduced by increasing
the amplification of the optical amplifier OA. As a matter of fact,
errors happen in any communication systems. In principle, there is
no limitation on the number $m$. By increasing the amplification,
one may reduce this ``quantum'' error rate to be well below other
error rates, such as errors related to the detector efficiency that
occurs in all telecommunication systems, and make this system as
reliable as any classical telecommunication system.

The time $\tau$ necessary for transmitting one bit of information
depends on the photon detecting processes at the sender and at the
receiver. But what is important is the fact that this time $\tau$
does not depend on the spatial separation between the sender and the
receiver. Thus if the distance between the sender and the receiver
is larger than $\tau c$, with $c$ the light velocity in vacuum, then
the signal transmitted from the sender to the receiver travels at a
superluminal speed. In this case, this telecommunication system
based on entangled photon pairs becomes a superluminal one.

The transmission of signals at a superluminal speed is in
contradiction to the relativistic causality. This telecommunication
system that could transmit information at superluminal speeds is
based on the polarization entangled photon pairs, therefore the
existence of the polarization entangled photon pairs is in direct
contradiction to the relativistic causality.

At this point, we are forced to make choice between the possibility
of producing quantum entanglement and the correctness of
relativistic causality. For most people, the breaking of the
relativistic causality is unacceptable. In my point of view, the
contradiction between the relativistic causality and existence of
quantum entanglement just implies the impossibility of generating
entangled quantum states. Many authors consider the photon pairs
emitted in a radiative atomic cascade of calcium \cite{asp} and
photon pairs generated from spontaneous parametric down-conversion
\cite{kwr} (SPDC) as polarization entangled photon pairs. But this
treatment is incorrect. As I pointed out in a recent article, the
photon pairs generated from SPDC are in an un-entangled quantum
state \cite{wang}
\begin{equation}\label{f_u}
    |\psi_u \rangle =
    \frac{1}{2}( b^\dag_{1} b^\dag_{1} +
    b^\dag_{2} b^\dag_{2})|0\rangle,
\end{equation}
with the positive frequency part of the vector potential in the
signal beam given by
\begin{equation}\label{a_s}
    \vec A^+_s = \frac{B}{\sqrt{2}}(b_1 \vec e_v - b_2 \vec e_h)
    e^{i\vec k_s \cdot \vec r} ,
\end{equation}
and in the idler beam given by
\begin{equation}\label{a_i}
    \vec A^+_i = \frac{B}{\sqrt{2}}(b_2 \vec e_v + b_1 \vec e_h)
    e^{i\vec k_i \cdot \vec r} ,
\end{equation}
where $B$ is a constant, $\vec e_v,\vec e_h$ are vectors of unit in
vertical and, respectively, horizontal directions, and $\vec k_s,
\vec k_i $ are wave vectors for signal beam and idler beam. We have
\begin{equation}\label{2f}
    \vec A_s^+ \vec A_s^+ |\psi_u \rangle \neq 0 \;\;
    \text{and} \;\;
    \vec A_i^+ \vec A_i^+ |\psi_u \rangle \neq 0 .
\end{equation}
That means for photon pairs in the quantum state $|\psi_u \rangle$,
beside the possibility of detecting one photon in each beam, both of
the photons can also be detected either in the signal beam or in the
idler beam at the same time. The probability of detecting one photon
in each beam is only 50\% for photon pairs in this quantum state. If
the photon pairs generated from SPDC are used in the above described
telecommunication system, then the correct information is
transmitted only if one and just one of the photon in a photon pair
is detected at the sender. But this happens with a probability equal
to 50\% only. In another word, 50\% of the signals received by the
receiver are erroneous. Because the correct information can be
either ``1'' or ``0'', one can guess the correct information with a
probability equal to 50\%, without receiving any signal. Therefore
an error portion of 50\% in signal readouts just means that no
information is transmitted, and superluminal telecommunication is
not realized. The same conclusion holds also for photon pairs
emitted in a radiative atomic cascade of calcium.

In conclusion, I have presented a schema for a telecommunication
system based on entangled photon pairs in which superluminal
transmission of information can be realized. In this
telecommunication system the existence of the polarization entangled
photon pairs is in direct contradiction to the relativistic
causality. In my opinion, this contradiction just implies the
impossibility of generating entangled quantum states.

\end{document}